# Specialisation and experience of research teams: Which matters more for the impact of their publications?

Emil Dolmer Alnor

Danish Centre for Studies in Research and Research Policy, Aarhus University, Denmark

*ea@ps.au.dk*

ORCID: 0000-0002-8536-9442

ABSTRACT: Scientists' topic choices strongly influence both individual careers and the advancement of the scientific frontier. While a sizeable body of literature shows that specialisation in a few topics benefits individual careers and fosters impactful research, the role of research teams and their experience have been largely overlooked. This paper introduces experience as a concept distinct from specialisation and shifts the level of analysis from the individual to the research team, reflecting the increasingly team-based nature of science. Using novel publication-level measures of team specialisation and team experience applied to nearly 1 million biomedical publications, the study finds that both are positively associated with citation impact. However, the correlation with citation impact is markedly stronger for team experience than for team specialisation. The study demonstrates how science can be examined at the team level and suggests that future research should pay more attention to studying experience.

## 1. INTRODUCTION

The development of the scientific frontier is largely determined by what topics scientists choose to pursue. At the same time, these choices strongly affect the success of individual scientists and the teams they compose. The evolution of topic choice among scientists and its consequences for individual careers have been widely studied (Abramo et al., 2019; de Rassenfosse et al., 2022; Chakraborty et al., 2015; Foster et al., 2015; Jamali et al., 2020; Leahey, 2007; Leahey et al., 2010; Zeng et al., 2019; Hill et al., 2025; Yu et al., 2021; Teodoridis et al., 2019; Heiberger et al., 2021). This literature focuses on the specialisation of individual researchers, how it benefits their careers, and how it affects the impact of their work. Specialisation can refer to the extent to which a researcher's body of work is concentrated within a narrow set of topics (Leahey, 2007, p. 538).

The present study introduces experience as a concept distinct from specialisation, defining the former as the extent to which a researcher has engaged with a topic. To illustrate the conceptual difference, consider a postdoc with five publications, all of which are in the topic 'Heart Transplantation', and a professor with 50 publications, half of which are in 'Heart Transplantation' and the other half spread across various other topics. The postdoc is more specialised in 'Heart Transplantation', but the professor has more experience in the topic. While the research cited above has extensively studied how individuals' specialisation shapes their careers and the impact of their research portfolios, the role of their experience has been overlooked.

Furthermore, the previous literature has ignored that individual researchers typically pool their expertise to conduct research in collaborative research teams. Scientific work is increasingly conducted by teams whose members contribute distinct skills and topic-specific knowledge (Wuchty et al., 2007; Jones, 2009). As a result, research publications typically reflect the combined capabilities of a research team rather than those of an individual (team member), and the influence of specialisation and experience on research outcomes cannot be fully understood by examining individual researchers in isolation. It is therefore particularly relevant to study specialisation and experience at the level of the research team.

Against this background, the purpose of the present study is to develop a measure of researcher experience, propose an approach to study experience and specialisation at the level of the research team, and examine how the extent of team specialisation and experience in the topics of their publications are associated with citation impact. Citation impact is widely studied in the literature cited above and is an important feature of research outcomes because it strongly affects the career prospects of researchers (Stephan et al., 2017). The sections below review the literature on how specialisation relates to the success of researchers and the impact of their publications, elaborate on the relevance of experience, identify knowledge gaps in the literature, and present the specific research question guiding this study.

### 1.1 Theoretical argument for specialisation and the relevance of experience

The theoretical arguments provided in the literature for how specialisation benefits research careers and increases the citation impact of publications can be categorised into a cognitive and a social mechanism. The cognitive mechanism is summarised by Leahey (2007, p. 540): "Specializing in one or a few subfields allows one to master the literature, to become familiar with important debates and gaps and recognize new developments." Sustained focus on a topic builds the knowledge and skills needed to conduct and publish research of high quality and relevance to that topic, which in turn increases citation impact (Tahamtan et al., 2016).

Regarding the social mechanism, de Rassenfosse et al. (2022) argue that specialisation is beneficial because it enables researchers to join the 'invisible colleges' surrounding knowledge production in specific topics (Crane, 1972), increasing researchers' reputation and visibility within these communities and ultimately leading to better opportunities for collaboration and more citations. Leahey (2007, p. 540) summarises this:

> By devoting a large portion of one's research to a limited number of areas, a scholar can come to know — and be known by — key players by attending smaller conferences and the relevant section meetings of larger conferences, reviewing peers' works in progress and manuscripts under review, and collaborating with researchers who have shared interests.

Here, we argue that experience is a relevant concept to study because the cognitive and social benefits do not stem from specialisation *per se*, but from the experience that specialisation enables. Given the complexity of modern science, specialisation may be a necessary condition for having the time required to develop deep expertise within a topic (Jones, 2009), but ultimately, it is not the decision to focus on a few topics, but rather the actual accumulated experience within those topics that fosters familiarity with their literatures and communities. Consider the cognitive mechanism: mastery of the theory and skills relevant to a scientific topic comes by sustained engagement with a topic, that is, with experience. It is likely easier to develop an understanding of what constitutes a meaningful scientific contribution in a field by concentrating one's efforts on that field, but ultimately, developing such understanding requires engagement with the field, not merely the decision to concentrate. The same holds for the social mechanism: a researcher's reputation within a community depends more directly on their accumulated activity in that community than their concentration of efforts in that community. Specialisation therefore primarily influences the citation impact of publications insofar as it facilitates the accumulation of experience.

Moreover, as a proxy for the activation of the social and cognitive mechanisms, specialisation has two shortcomings that experience does not. The first shortcoming is that equal levels of specialisation can hide substantial variation in skills and reputation. A researcher with three years of work concentrated entirely on a single topic and a researcher with twenty years of the same concentration are equally specialised yet differ substantially in their familiarity with the field and their standing within it. More generally, specialisation is insensitive to the volume of engagement: it measures the *proportion* of a researcher's work in a topic, not the *amount*.

The second shortcoming is that specialisation decreases when researchers work outside the focal topic, whereas the factors argued to matter for citation impact — skills and reputation — do not. Consider the two researchers from the introductory example: the postdoc with all five publications in 'Heart Transplantation' is fully specialised, while the professor with 25 publications in 'Heart Transplantation' and 25 in other topics is less specialised. Yet the professor's skills and reputation within 'Heart Transplantation' are not diminished by those other 25 publications — if anything, breadth may enrich them. Specialisation thus declines with engagement in other topics even when that engagement does not reduce, and may even support, researchers' ability to produce high-impact work in the focal area.

Experience, or the extent of a researcher's engagement with a topic, directly alleviates both shortcomings. It is sensitive to the volume of prior work in a topic and it is not reduced by engagement with other topics. Experience therefore tracks the activation of the cognitive and social mechanisms more directly than specialisation does, and we expect it to be a stronger predictor of citation impact.

### 1.2 Empirical findings of previous studies

We now review the empirical findings of previous studies. The review pays particular attention to their methodological approaches to motivate the development of the experience measure and the method of how to study it at the team level.

Leahey (2007) analyses specialisation among sociologists. The degree of specialisation is measured by dividing the number of unique keywords in a sociologist's publication portfolio by the number of publications — higher values indicate greater specialisation (Leahey, 2007, p. 543). She finds that specialisation is associated with increased productivity (number of publications) and earnings. In a similar study, Leahey et al. (2010) examine how specialisation affects tenure among sociologists and — contrary to the general literature — find that specialisation is negatively related to promotion prospects. In contrast, Heiberger et al. (2021) find that sociologists who continuously publish in topics similar to their dissertation are more likely to become supervisors; that is, specialising in the topic of one's dissertation is beneficial. Hill et al. (2025) also find that maintaining topical stability is beneficial. They operationalise topic switching as a continuous measure based on the cosine similarity between a focal paper's referenced journals and those in the author's previous work. They find that researchers who switch topics receive fewer citations, indicating that specialisation increases impact.

Jamali et al. (2020) study diversification (the opposite of specialisation) in the publication portfolios of Australian professors from Physics, Chemistry, and Biology. They use the Australian Fields of Research (FoR) to create four measures of the extent of diversification: 1) number of unique FoR divided by number of publications; 2) 1 minus modal percentage (share of publications belonging to the most typical FoR); 3) share of publications coded as Multidisciplinary; and 4) variance of FoR codes, which they argue is a valid measure of diversification because FoR codes are integers with neighbouring disciplines having similar values. They find that diversification is negatively associated with both citation impact and productivity. Abramo et al. (2019) apply an approach similar to the '1 minus modal percentage' approach and find that the most cited publication in Italian professors' publication portfolio typically belongs to the field in which they publish the most. In their words, specialised publications receive more citations than diversified publications.

Yu et al. (2021) analyse changes in research direction. For each researcher, they construct two vectors of keywords from the Physics and Astronomy Classification Scheme, representing early- and late-career publications, respectively. Using cosine similarity, they find that researchers who make larger changes in research direction increase their citation impact more. Since direction change is somewhat counter to specialisation, their finding may indicate that specialisation decreases citation impact. However, their findings align with those of de Rassenfosse et al. (2022) and Chakraborty et al. (2015), who distinguish between period-specific and career-level specialisation. In Chakraborty et al. (2015), computer science publications are assigned a field using the Microsoft Academic Search Engine, and specialisation is then proportional to the number of distinct fields in a researcher's publication portfolio relative to their number of publications. They find that the researchers with the largest citation impact work on diverse topics throughout their entire career but specialise in one/two fields at any given time. De Rassenfosse et al. (2022) also find that period-specific specialisation is beneficial, whereas career-level specialisation is not. In their operationalisation, researchers are specialised when they publish a high share of papers indexed with a specific Medical Subject Heading (MeSH) term compared to the global share of papers indexed with that MeSH term. They divide biomedical researchers' publication portfolios into three-year windows and find that specialisation within windows increases citation impact, but also that citation impact increases with changes in content between windows.

Zeng et al. (2019) also consider career-timing aspects, although differently. They construct 'for each scientist a co-citing network (CCN), in which each node is a paper authored by this scientist and two papers are linked if they share at least one reference'. (Zeng et al., 2019, p. 2). They then apply community detection to these networks to detect clusters. A topic switch is defined as a researcher publishing two time-adjacent papers belonging to different clusters. Their results show that researchers who frequently switch topics receive fewer citations; however, the most productive researchers rarely switch topics early in their careers but often switch topics later in their career.

Finally, Teodoridis et al. (2019) also consider the timing of specialisation but focus on the knowledge environment surrounding the scientist. They find that specialist mathematicians (i.e., those with publications in only one Mathematics Subject Classification) had greater impact and productivity after the collapse of the Soviet Union than generalist mathematicians but that the opposite was true before the collapse. They argue that specialisation is beneficial in fast-changing knowledge environments, where new knowledge quickly becomes accessible, as was the case after the collapse of the Soviet Union, but that being a generalist is more beneficial in slow-paced knowledge environments.

### 1.3 Contributions of the present study

The studies reviewed above provide considerable insight into how and when specialisation benefits individual careers and publications. Although the body of work is methodologically diverse, it also has similarities that create blind spots and leave important questions unanswered. First, the level of analysis is the individual researcher in every study. Whether using keywords, fields, or referenced journals, measurements are always based on the publication portfolio of the individual scientist, and specialisation levels are assigned to individuals. As stated in the introduction, teams have become increasingly responsible for the processes and products of science, to the extent that single-authored publications are uncommon. It is therefore crucial to study specialisation at the level of the research team. Second, how are individual publications affected by the specialisation levels of their authors? Apart from the studies by Abramo et al. (2019) and Hill et al. (2025), the literature has only examined the consequences of specialisation at aggregate levels, such as how the entire publication portfolio of a researcher is influenced by the extent to which their body of work is concentrated within a narrow set of topics. However, specialisation, if it matters, must ultimately exert its influence through individual publications. Third, what is the influence of experience? Although experience and specialisation may appear similar at first glance, the introduction has argued that they are conceptually distinct and that experience is more directly linked to citation impact than specialisation. Previous studies have focused exclusively on specialisation, leaving the role of experience largely unexplored. Against this background, the present study investigates the following research question:

> How are the specialisation and experience of a research team in a publication's topic associated with the publication's citation impact?

The empirical analysis focuses on the biomedical sciences, pragmatically defined as research indexed in PubMed. This field constitutes one of the largest domains of contemporary science, accounting for a substantial share of global research activity and public investment. Moreover, the biomedical sciences are characterised by pronounced team science dynamics, making them particularly well suited to address the research question.

The rest of the paper proceeds as follows. Section two expands on the conceptual and mathematical differences between specialisation and experience and develops two novel measures to capture their extent at the research team level with respect to an individual publication. Section three describes the dataset and variables used in a set of regression models with the likelihood of being a top 5% cited publication as the dependent variable. Section four analyses the correlation between specialisation/experience and citation impact in a large number of publications using a cross-sectional (n = 992,519), fixed effects (n = 31,683), and two-way fixed effects design (n = 9,588). The results show that both specialisation and experience are substantively correlated with citation impact, but the correlation for experience is markedly larger. The fifth and final section discusses implications for research and policy, outlines the limitations of the study, and suggests directions for future research.

## 2. MEASURING TEAM SPECIALISATION AND EXPERIENCE IN A PUBLICATION'S TOPIC

We adapt the definition by Leahey (2007) and define team specialisation in a publication's topic as the extent to which the team's body of work is concentrated within the topic of the publication. Team experience in a publication's topic is the extent to which the team has engaged with the topic of the publication. Despite their surface similarity, the two concepts differ in two important respects. First, specialisation is a bounded concept: a team cannot be more specialised than fully specialised. This occurs when the topics of their previous body of work are perfectly aligned with the topics of the publication. In contrast, experience is an unbounded concept: even when perfectly aligned, a team can gain further experience in a publication's topic, if they do more research in that topic. Second, research outside the focal publication's topic reduces specialisation but not experience. For example, consider a publication on "Heart diseases" and two teams: Team A has worked on "Heart diseases" for 10 years, while Team B has 10 years of work on "Heart diseases" and 10 years of work on "Vascular Diseases". Team A is more specialised in the topic of the focal publication, but Team B's experience in "Heart diseases" is not diminished by their additional work on "Vascular Diseases".

### 2.1 Representing research team publication records and focal publication topics as MeSH-based vectors

To properly account for the multifaceted nature of science — teams work on various topics throughout their careers, and a publication may simultaneously concern several topics (Held, 2022) — we represent publications and teams' previous work using vectors, where each dimension corresponds to a topic and its value reflects the importance of that topic. The controlled vocabulary MeSH provides a natural basis for these vectors. MeSH is used to index publications in PubMed. It contains more than 30,000 terms that describe scientific concepts and that are hierarchically organised; for example, "Heart Arrest" is a subterm of "Heart Diseases" (U.S. National Library of Medicine, 2024; Alnor, 2026). The publications used in this analysis are on average indexed with 12 MeSH terms, with an average of four terms designated as "major", which reflect the primary topics of the publication. The remaining "minor" MeSH terms reflect concepts of lesser importance to the publication.

Alnor (2026) tests different approaches to measuring the relatedness between scientific publications using controlled vocabularies and finds that accuracy is increased by adding weights to major or rare terms and by accounting for the similarity between non-matching terms. Since measuring publication relatedness is analogous to measuring specialisation and experience, we implement these findings in the operationalisation of specialisation and experience below.

We first describe how a publication enters the publication record vectors of its authors. Represent publication $p$ with the vector $\mathbf{p} = [p_1, \ldots, p_n]$, where $n$ denotes the number of unique MeSH terms. Let $p_i = 1$ if MeSH term $i$ in $p$ is a minor term and $p_i = 3$ if MeSH term $i$ is a major term in $p$. Major terms are weighted 3 because of the results in Alnor (2026). We normalise $\mathbf{p}$, so its elements sum to 1, specifically, $p'_i = p_i / \sum_{j=1}^{n} p_j$. An individual researcher's publication record vector in year $y$, $r^y$, is the sum of the publication vectors for publications prior to $y$. The team's publication record vector is the sum of its members' publication record vectors, $t_i^y = \sum_{i=1}^{n} r_i^y$, where $n$ is the number of team members.

To add a weight to rare terms we introduce the concept of information content (IC) (Ahlgren et al., 2020). The IC of a MeSH term is inversely proportional to the probability of encountering the term or its descendants; that is, rare terms have a high IC. Formally, let $\#(m_i)$ denote the frequency of MeSH terms *i* in the dataset, let $D_i$ denote the set of terms that are descendants to $m_i$ either directly or indirectly, and let *n* denote the number of unique MeSH terms. Then, the IC of $m_i$ is

$$\mathrm{IC}(m_i) = -\log\left(\frac{\#(m_i) + \sum_{d \in D_i} \#(d)}{\sum_{k=1}^{n} \#(m_k) + \sum_{d \in D_k} \#(d)}\right) \tag{1}$$

We can now represent the research team and the focal publication as two vectors, **a** and **b**, respectively, where $a_i = t_i^y \times \mathrm{IC}(m_i)$ and

$$b_i = \begin{cases} 0 \text{ if } p \text{ is not indexed with } m_i \\ \mathrm{IC}(m_i) \text{ if } m_i \text{ is a minor term in } p \\ \mathrm{IC}(m_i) \times 3 \text{ if } m_i \text{ is a major term in } p \end{cases} \tag{2}$$

### 2.2 Using research team and focal publication vectors to measure specialisation and experience

With the research team and focal publication represented as vectors, two vector operations naturally correspond to the theoretical concepts. Specialisation is captured by soft cosine similarity, which measures the angle between the two vectors. Experience is captured by scalar projection, which measures the magnitude of the team vector in the direction of the publication vector.

Soft cosine similarity is a modification of cosine similarity that accounts for the similarity between non-matching terms (Sidorov et al., 2014). To quantify term similarity we follow the most accurate method identified in Alnor (2026) and create a weighted undirected graph, where the nodes are MeSH terms, edges exist between parent–child MeSH terms (e.g., between “Heart Diseases” and “Heart Arrest”), and weights are defined by the difference in IC between parent and child. The distance between two MeSH terms in the graph, $\mathrm{dist}(i, j)$, is the sum of weights on the shortest path connecting them. Distance is converted to similarity using $\mathrm{sim}(i, j) = \frac{1}{\exp(\mathrm{dist}(i,j))}$.

Now with **S** being the term-similarity matrix, where $s_{ij}$ measures the similarity between terms *i* and *j*, we define specialisation as the soft cosine between the team vector, **a**, and the publication vector, **b**:

$$Specialisation = \frac{\sum\sum_{i,j=1}^{n} s_{ij} a_i b_j}{\sqrt{\sum\sum_{i,j=1}^{n} s_{ij} a_i a_j}\sqrt{\sum\sum_{i,j=1}^{n} s_{ij} b_i b_j}} = \frac{\mathbf{a}^T \mathbf{S} \mathbf{b}}{\sqrt{\mathbf{a}^T \mathbf{S} \mathbf{a}}\sqrt{\mathbf{b}^T \mathbf{S} \mathbf{b}}} \tag{3}$$

Experience is the scalar projection of the team vector onto the publication vector, accounting for similarities between terms[1]

$$Experience = \frac{\mathbf{a}^T\mathbf{S}\mathbf{b}}{\sqrt{\mathbf{b}^T\mathbf{S}\mathbf{b}}} \tag{4}$$

Just as the concepts are similar, so are their operationalisations. The difference is the term $\sqrt{\mathbf{a}^T\mathbf{S}\mathbf{a}}$, which is present in the specialisation formula but not the experience formula. Because this term is in the denominator of the specialisation formula and increases with the size of team's publication record vector it accounts for the conceptual differences between specialisation and experience: it imposes the upper bound on specialisation and reduces specialisation when teams have publications in MeSH terms that are not in the focal publication.

Specialisation ranges from 0 (no specialisation: no publications in any MeSH terms of the publication) to 1 (full specialisation: only publications in the MeSH terms of the publication). Experience is a positive continuous variable with higher values indicating more experience. It is right-skewed, and we expect diminishing returns to experience with respect to citation impact. To account for this in the regression analysis, we log-transform experience. Figure 1 shows the distribution of specialisation and the original and log-transformed experience variables.

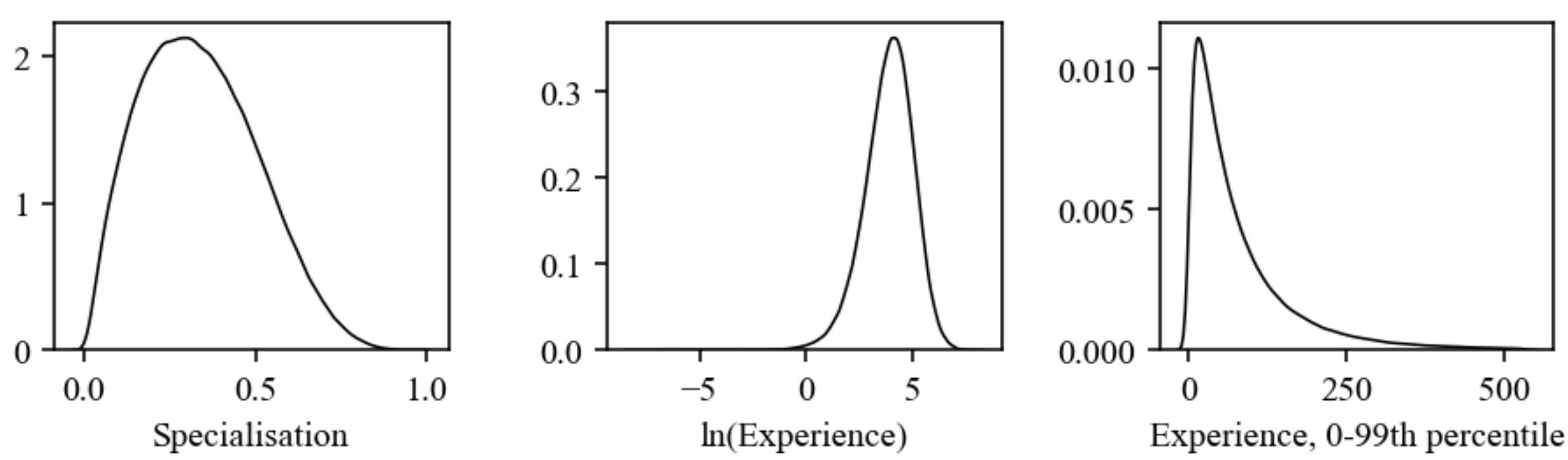

**Fig. 1** Density plots of team specialisation, natural log-transformed team experience, and (untransformed) team experience truncated at the 99th percentile.

---

[1] In the simplistic case of no similarity between terms ($s_{ij} = 0$ for $i \neq j$) soft cosine similarity reduces to standard cosine similarity:

$$\frac{\mathbf{a}^{\mathrm{T}}\mathbf{S}\mathbf{b}}{\sqrt{\mathbf{a}^{\mathrm{T}}\mathbf{S}\mathbf{a}}\sqrt{\mathbf{b}^{\mathrm{T}}\mathbf{S}\mathbf{b}}} = \frac{\mathbf{a}\cdot\mathbf{b}}{||\mathbf{a}||\,||\mathbf{b}||}, \mathbf{S} = \mathbf{I}$$

And experience is a standard scalar projection:

$$\frac{\mathbf{a}^T\mathbf{S}\mathbf{b}}{\sqrt{\mathbf{b}^T\mathbf{S}\mathbf{b}}} = \frac{\mathbf{a}\cdot\mathbf{b}}{||\mathbf{b}||}, \mathbf{S} = \mathbf{I}$$

### 2.3 Summary

Specialisation quantifies the extent to which the terms in the team's publication record and the terms of the publication are aligned. Experience quantifies the extent to which the team has published in the terms of the publication.

Like all previous studies reviewed in Section 1.2, we rely on publication records in our operationalisations. The advantage of publication records is that they allow large-scale analysis — the present study analyses nearly 1 million focal publications. Their drawback is that researchers may not have made substantial work on each publication in their portfolios, particularly due to the prevalence of honorary authorships. While we do not solve this problem, we take several steps to improve measurement validity relative to previous studies: using vectors rather than one-dimensional subject categories, adding weights to major and rare terms, and accounting for term similarity. We return to the issue of honorary authorships in the discussion.

Neither specialisation nor experience are inherent qualities of a team but are measured with respect to individual publications. A team's extent of specialisation and experience varies by publication, and they can publish two papers in the same year and be experts or specialists in one but not in the other. This means that the levels of experience and specialisation can vary while all factors that do not vary within teams and years can be held constant, thereby minimizing the influence of other team factors related to citation impact such as team seniority or their number of previous publications.

## 3. METHODS

This section describes the data and the dependent and control variables used in a series of logit regressions modelling publication status as top 5% most cited.

### 3.1 Data and descriptive statistics

We use the 'Author-ity 2009 – PubMed author name disambiguated dataset' (Torvik & Smalheiser, 2018) to assign publications to authors. The experience and specialisation variables may become unstable in publications with many authors, so we exclude the 1% of publications with 12 or more authors. Since our purpose is to shift the level of analysis to the team, we exclude single-authored publications[2]. The remaining data are matched to the CWTS WoS database to obtain MeSH and citation impact data. We only include publications where each author has at least three previous publications with MeSH terms. After applying these filters, the dataset consists of 992,519 publications from 2000 to 2008. Table 1 shows descriptive statistics for these publications.

[2] The regression tables in the appendix shows that the results are robust to including single-authored publications (n = 1,198,575) and restricting the sample to publications with three or more authors (n = 732,328).

**Table 1** Descriptive statistics for main variables.

| Variable | Mean | Standard deviation | Min | 10% | 50% | 90% | Max |
|---|---|---|---|---|---|---|---|
| Specialisation | 0.339 | 0.169 | 0 | 0.124 | 0.327 | 0.572 | 0.98 |
| ln(Experience) | 3.9 | 1.17 | -8.27 | 2.39 | 3.99 | 5.3 | 8.14 |
| Top 5% | 0.068 | 0.252 | | | | | |
| #Authors | 4.1 | 2.1 | 2 | 2 | 4 | 7 | 11 |
| ln(#Previous publications) | 5 | 1 | 1.8 | 3.6 | 5.1 | 6.3 | 8.3 |
| Review | 0.115 | 0.319 | | | | | |

n = 992,519. Review is coded 'review = 1' and 'research article = 0'.

### 3.2 Citation impact and dichotomising the top 5% indicator

To measure citation impact, we consider whether a publication is among the top 5% most cited publications 10 years after publication when compared to publications within the same field and year (top 5%). Fields are defined using the CWTS micro-clusters.

Most publications are either top 5% or not. However, because many publications within a field-year have identical citation counts, it is generally not possible to have exactly 5% publications in the top 5% most cited; that is, the proportion will be lower or higher than 5%. Waltman and Schreiber (2012) address this by assigning publications with a tie citation score to the top 5% with a fraction. While this solves the tie issue in portfolios of publications, it is still an open question how to handle individual publications (Waltman & Schreiber, 2012, p. 378), which must be done in the current analysis. A rule of thumb, which some analysts may use, is to assign a publication to the top 5% when its fraction is 0.5 or greater. This is problematic since it systematically causes fields to have different proportions of top 5% publications, which is exactly the problem that Waltman and Schreiber (2012) try to solve.[3]

Instead, we use the original continuous top 5% indicator to construct a strictly binary indicator through probabilistic sampling. In the new binary top 5% indicator, each publication is assigned to the top 5% based on a Bernoulli trial using its fractional score as the success probability. This means that publications with a fractional score of 1 are top 5% (with 100% probability), publications with fraction of 0 are not top 5%, while publications with a fractional value between 0 and 1—because of a tie citation score— are assigned randomly to either top 5% or non-top 5% according to their probability. This approach has the desirable statistical property that the expected proportion of top 5% publications equals the mean value of the original (non-binary) top 5% indicator.[4]

[3] To exemplify, consider three sets of 100 publications from different field-years. All publications have a tie citation score. In set A, each publication has a fractional top 5% inclusion of 0.49; in set B, 0.51; and in set C, 0.95. Using the rule of thumb, no publication in set A will be in the top 5%, while all publications in sets B and C will. The rule of thumb clearly exaggerates the difference in citation impact between sets A and B while understating the difference between sets B and C.

### 3.3 Control variables

The regression models control for publication type (review = 1, research article = 0), number of authors in the research team, and total number of previous publications by the authors. To account for diminishing returns to citation impact, the number of previous publications is log transformed.

### 3.4 Strategy of analysis

We estimate a series of logit models with top 5% as the dependent variable, sequentially adding stricter controls. The first five models include a large number of publications but have relatively low internal validity because they only control for select observable features. In models VI and VII, we utilise that both experience and specialisation are defined with respect to a publication and add team-fixed effects. A team can be experts in the topic of one of its publications but not another, and this allows comparison of top 5% and non-top 5% publications by the same team to assess whether the top 5% publications have higher specialisation/experience levels. These models have substantially fewer observations as they only include publications by teams (consisting of the exact same group of authors) that have both top 5% and non-top 5% publications but provide stronger internal validity by controlling for unobserved heterogeneity at the team level. However, they do not control for variables that vary within teams across time (e.g., team dynamics). Therefore, Models VIII and IX employ team *and* year fixed effects. This restricts the sample to publications by teams that have both a top 5% and a non-top 5% publication within the same year.

## 4. RESULTS

This section analyses the extent to which specialisation and experience are correlated with top 5% likelihood. As an initial analysis, Table 2 compares the mean specialisation and experience of top 5% and non-top 5% publications. The results show that both variables have higher mean values for top 5% publications. This provides initial evidence that specialisation and experience are positively associated with the probability of a publication being highly cited.

---

[4] It also has the undesirable statistical property of introducing some randomness into the dependent variable (6,547 / 0.7% of 992,519 publications used in models I-V have a tie citation score). The appendix tests the robustness of the results to alternative realizations of the Bernoulli trial by re-estimating the models across 10,000 independent trials. The results show that the variation in the coefficient estimates does not affect the conclusions.

**Table 2** Mean of specialisation and experience by top 5% status.

| Top 5% | Specialisation | ln(Experience) |
|---|---|---|
| No | 0.336 | 3.88 |
| Yes | 0.378 | 4.19 |

The results in Table 3 corroborate these findings, showing that specialisation and experience are positively correlated with top 5% likelihood across all models. This means that publications by teams with higher specialisation or experience are more likely to be top 5% than publications by teams with lower specialisation or experience. Since the number of observations in all models is very high, we omit significance testing to avoid inducing a false certainty in the results and instead focus on the substantive magnitude of the coefficients.

Models I and II regress top 5% likelihood on only specialisation and experience, respectively. Model III includes only control variables, while models IV and V include both specialisation and experience, respectively, and control variables. To facilitate interpretation of the coefficient sizes, Figure 2 shows the predicted probability of top 5% at different specialisation and experience levels, based on the estimates from models IV and V. Control variables are set at their mean/mode. Publications by teams with specialisation one standard deviation above the mean have 2.2 percentage points (55%) higher probability of being top 5% than publications by teams one standard deviation below the mean. For experience, the corresponding difference is 4.3 percentage points (133%). While the differences in predicted top 5% probability are large for both specialisation and experience, the difference is more than double for experience compared to specialisation $\left[\frac{133}{55} \approx 2.41\right]$. These results indicate that experience is a substantially better predictor of top 5% likelihood than specialisation.

**Table 3** Logit and conditional logit regressions of top 5%.

| | Model | | | | | | | | |
|---|---|---|---|---|---|---|---|---|---|
| | Logit | | | | | Conditional logit (Team fixed effects) | | Conditional logit (Team and year fixed effects) | |
| | I | II | III | IV | V | VI | VII | VIII | IX |
| Specialisation | 1.420<br>(0.023) | | | 1.371<br>(0.023) | | 0.903<br>(0.103) | | 1.198<br>(0.188) | |
| ln(Experience) | | 0.243<br>(0.004) | | | 0.381<br>(0.006) | | 0.230<br>(0.030) | | 0.315<br>(0.053) |
| Review | | | 1.508<br>(0.010) | 1.488<br>(0.010) | 1.499<br>(0.010) | 1.237<br>(0.036) | 1.239<br>(0.036) | 1.093<br>(0.066) | 1.098<br>(0.066) |
| ln(#Previous publications) | | | 0.166<br>(0.005) | 0.177<br>(0.005) | -0.165<br>(0.007) | -0.251<br>(0.065) | -0.459<br>(0.073) | | |
| #Authors | | | 0.001<br>(0.003) | -0.006[*]<br>(0.003) | 0.001<br>(0.003) | | | | |
| Constant | -3.125<br>(0.010) | -3.598<br>(0.016) | -3.760<br>(0.022) | -4.279<br>(0.024) | -3.621<br>(0.022) | | | | |
| n | 992,519 | 992,519 | 992,519 | 992,519 | 992,519 | 31,683 | 31,683 | 9,588 | 9,588 |
| n top 5% | 67,476 | 67,476 | 67,476 | 67,476 | 67,476 | 12,299 | 12,299 | 4,365 | 4,365 |
| Log likelihood | -244,665 | -244,223 | -234,008 | -232,320 | -231,852 | -10,370 | -10,378 | -3,185 | -3,188 |

Note: Log-odds coefficients (standard errors). Review is coded 'review = 1' and 'research article = 0'.

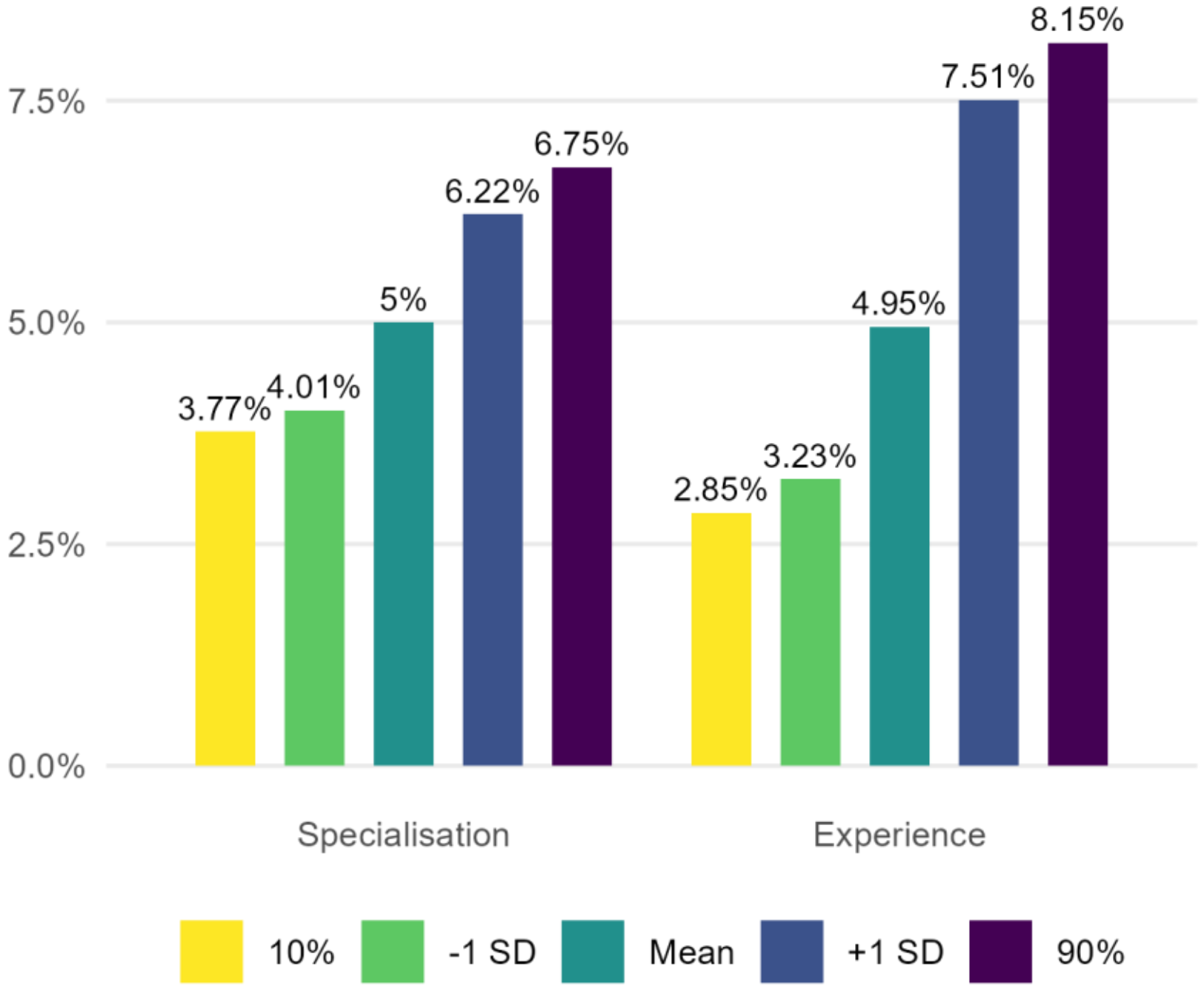

**Fig. 2** Predicted probability of top 5% by level of specialisation and experience**.** Research articles by teams with mean number of authors and ln(previous publications). Probabilities are based on estimates from model IV (specialisation) and V (experience) in Table 3.

Models VI and VII employ conditional logit regression with team fixed effects, while Models VIII and IX use team-year fixed effects. Specialisation and experience are positively and substantively correlated with top 5% likelihood in all models. Figure 3 shows odds ratios of top 5% for selected comparisons in specialisation and experience levels. Consider two publications, A and B, published by the same research team in the same year. If the team's specialisation score is one standard deviation above the mean for publication A and one standard deviation below for publication B, publication A has 50% higher odds of being top 5% than publication B $\left[\exp\left(1.198 \times (0.508 - 0.171)\right) \approx 1.50\right]$[5]. The corresponding difference for experience is 109% — roughly double that of specialisation $\left[\exp\left(0.315 \times (5.07 - 2.73)\right) \approx 2.09\right]$. Across all comparisons in Figure 3, both specialisation and experience show a substantial association with the odds of top 5%, but experience shows a markedly stronger association.

[5] For simplicity, we display rounded coefficients, percentiles, and differences in odds, but calculations are done using non-rounded numbers.

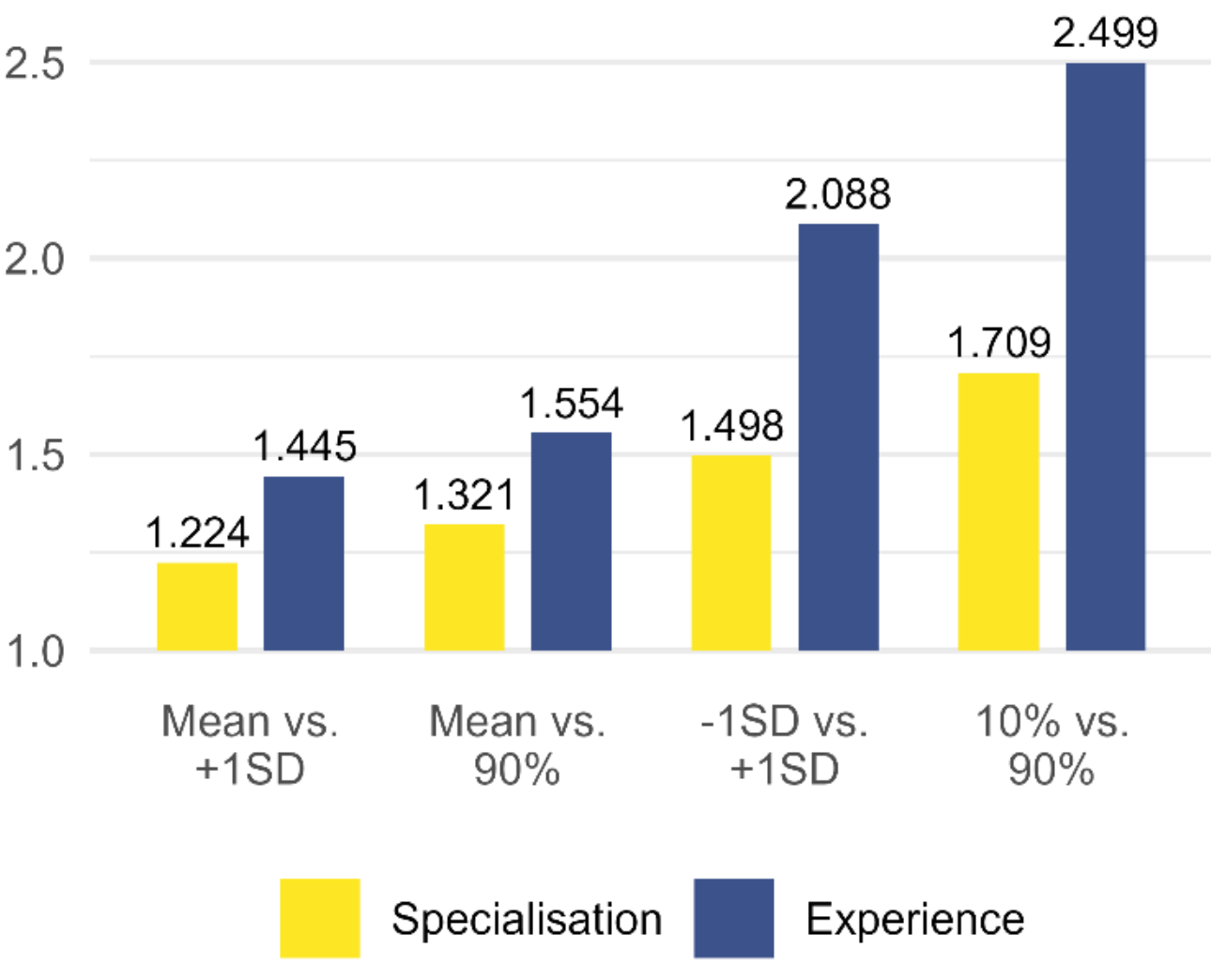

**Fig. 3** Odds ratios of top5% for selected comparisons in specialisation and experience levels. Odds ratios represent the odds of top 5% for the bottom level relative to the top level, e.g., odds of top5% with experience one standard deviation above vs. below the mean. Odds ratio is derived using $\exp(\beta \times (x_1 - x_2))$, where $\beta$ is the coefficient from model VIII (specialisation) or IX (experience) in Table 3, and $x_1$ and $x_2$ are the lower and upper comparison values, respectively.

The appendix shows the results of three robustness analyses. First, the models are re-estimated when restricting the publication sample to publications with at least three authors and when including single-authored publications. Across both alternative samples, the results are qualitatively similar to those in Table 3. Second, the models are re-estimated with alternative definitions of the team publication record vector. Recall that this vector was the sum of each individual team member's publication record vector, $t_i^y = \sum_{i=1}^{n} r_i^y$, with $n$ being the number of authors. In the appendix, the results are tested against two alternatives: defining the team publication record vector as the max of any of its team members' publication record vectors, $t_i'^y = \max(r_i^y, \dots, r_n^y)$, and as the mean, $t_i^{*y} = \frac{1}{n}\sum_{i=1}^{n} r_i^y$, respectively. Across both specifications, the results remain qualitatively similar to those in the main text. Third, the models are re-estimated across 10,000 independent Bernoulli trials used to dichotomize the top 5% indicator. The variation of the coefficient estimates does not alter the conclusions, meaning that the results are robust to the randomness introduced by the Bernoulli trials.

## 5. DISCUSSION AND CONCLUSION

This paper has found that both team specialisation and experience are positively and substantively correlated with citation impact in a large sample of biomedical publications (n = 992,519) with experience showing a markedly stronger correlation than specialisation. Analysis using team and year fixed effects on a smaller sample (n = 9,588) found the same results. The results demonstrate the importance of experience and the feasibility of analysing research teams rather than individuals. Below, we discuss implications of the results and outline limitations of the study and avenues for future research.

### 5.1 Implications for research

The sizeable body of existing literature on specialisation has focused on the individual researcher and has not adequately accounted for the fact that research more often than not is conducted in teams (Abramo et al., 2019; de Rassenfosse et al., 2022; Chakraborty et al., 2015; Jamali et al., 2020; Leahey, 2007; Leahey et al., 2010; Zeng et al., 2019; Hill et al., 2025; Yu et al. 2021; Teodoridis et al., 2019; Heiberger et al., 2021). This study has shown one approach to shift the level of analysis to the research team when examining specialisation and experience. The results show that the research team is highly relevant to study as their levels of specialisation and experience are substantively associated with the citation impact of their publications. This suggests that future research may benefit from a stronger focus on the team dimension of specialisation and experience.

Beyond neglecting the research team, the existing literature has also overlooked the potential role of experience. Theoretically, specialisation has two shortcomings in influencing citation impact via the cognitive and social mechanisms posited in the literature: it measures *proportion* rather than *volume* of engagement with a topic and declines with work outside the focal topic even though researchers' skills and reputation do not. We argued that experience more directly captures the accumulated engagement with a topic that activates the cognitive and social mechanisms. The empirical results are consistent with this argument: across all model specifications experience is more strongly associated with citation impact than specialisation is. More specifically, the predicted probability and the odds of making a top 5% cited publication change approximately twice as much between low- and high-experience teams than between low- and high-specialisation teams. Therefore, future research may benefit from placing greater emphasis on experience as it is currently understudied and appears to play a more significant role in explaining citation impact.

### 5.2 Implications for policy

One of the most significant research policy trends over recent decades has been an increasing amount of targeted mode and mission-oriented funding, where research funds are earmarked for prespecified topics or mission objectives (Gläser & Laudel, 2016; Whitley et al., 2018). Although researchers have options to resist external pressures to change their topical focus (Gläser, 2019), some studies suggest that targeted funding can cause researchers to temporarily shift away from their familiar areas (Myers, 2020; Madsen & Nielsen, 2024; Mancuso & Broström, 2026). The results of this study suggest that such shifts could result in less impactful research. These findings therefore support Gläser's (2019, p.

442) conclusion that the most effective way to promote research in specific topics is to provide resources to researchers already working in those areas.

### 5.3 Limitations and future research

Like all previous studies that use publication records to measure specialisation, the measurement validity of the present study is limited by the fact that not every author of a publication has necessarily collaborated closely or made substantial contributions. This issue is particularly problematic due to the prevalence of honorary authorships, which potentially distorts the measurement of specialisation and experience. While honorary authorships can increase citation impact through the social mechanism — by increasing honorary authors' exposure within a scientific community — they do not reflect the kind of experience that develops the ability to produce relevant, high-quality research. Data on contribution statements or alternative sources that better capture the structure and intensity of research collaboration are not available at the same scale as the data used in the present study but could be incorporated by future research to address the measurement validity issues that arise when authorship is decoupled from substantial contribution.

A second limitation of the study concerns its generalisability. The large number of publications included in the analysis provides generalisability to the biomedical sciences, but it remains an open question how the results generalise to other scientific fields. This could be investigated by future studies using other controlled vocabularies, such as the IEEE Thesaurus, the Physics Subject Headings (PhySH), or the ACM Computing Classification System (CCS).

Third, the aggregation from individual publication records to team publication records is a simplified representation of how team members actually complement each other. Although the results are robust to alternative aggregation rules, future research could theorise more on how skill diversity and complementarity in research teams relate to research outcomes.

## ACKNOWLEDGEMENTS

The author thanks Jens Peter Andersen, Vincent Traag, Maria-Theresa Norn, and participants at seminars at CFA, Aarhus University, and CWTS, Leiden University, for their comments and suggestions.

## COMPETING INTERESTS

The author has no competing interests.

## FUNDING INFORMATION

This research received no external funding.

## DATA AVAILABILITY

The 'Author-ity 2009 – PubMed author name disambiguated dataset' used to assign authors to publications is publicly available (Torvik & Smalheiser, 2018). The CWTS-WoS data with information on citation impact and MeSH is not publicly available. The Python and R code used to process and analyse the data is available at https://github.com/EDAlnor/spex.

## GENERATIVE AI USE

Claude and ChatGPT were used to assist with language editing in parts of the manuscript.

## REFERENCES

Abramo, G., D'Angelo, C. A., & Di Costa, F. (2019). Diversification versus specialisation in scientific research: Which strategy pays off? *Technovation*, *82-83*, 51-57. https://doi.org/10.1016/j.technovation.2018.06.010.

Ahlgren, P., Chen, Y., Colliander, C., & van Eck, N. J. (2020). Enhancing direct citations: A comparison of relatedness measures for community detection in a large set of PubMed publications. *Quantitative Science Studies*, *1*(2), 714-729. https://doi.org/10.1162/qss_a_00027.

Alnor, E. D. (2026). Measuring the relatedness between scientific publications using controlled vocabularies. https://doi.org/10.48550/arXiv.2602.14755.

Chakraborty, T., Tammana, V., Ganguly, N., & Mukherjee, A. (2015). Understanding and modeling diverse scientific careers of researchers. *Journal of Informetrics*, *9*(1), 69-78. https://doi.org/https://doi.org/10.1016/j.joi.2014.11.008

Crane, D. (1972). *Invisible colleges: diffusion of knowledge in scientific communities*. University of Chicago Press.

Foster, J. G., Rzhetsky, A., & Evans, J. A. (2015). Tradition and innovation in scientists' research strategies. *American sociological review*, *80*(5), 875-908. https://doi.org/10.1177/000312241560161

Gläser, J. (2019). How can governance change research content? Linking science policy studies to the sociology of science. In *Handbook on science and public policy* (pp. 419-447). Edward Elgar Publishing.

Gläser, J., & Laudel, G. (2016). Governing science: How science policy shapes research content. *European Journal of sociology*, *57*(1), 117-168.

Heiberger, R. H., Munoz-Najar Galvez, S., & McFarland, D. A. (2021). Facets of specialisation and its relation to career success: An analysis of US Sociology, 1980 to 2015. *American sociological review*, *86*(6), 1164-1192. https://doi.org/10.1177/0003122421105

Held, M. (2022). Know thy tools! Limits of popular algorithms used for topic reconstruction. *Quantitative Science Studies*, 1-25. https://doi.org/10.1162/qss_a_00217

Hill, R., Yin, Y., Stein, C., Wang, X., Wang, D., & Jones, B. F. (2025). The pivot penalty in research. *Nature*. https://doi.org/10.1038/s41586-025-09048-1

Jamali, H. R., Abbasi, A., & Bornmann, L. (2020). Research diversification and its relationship with publication counts and impact: A case study based on Australian professors. *Journal of Information Science*, *46*(1), 131-144. https://doi.org/10.1177/0165551519837191

Jones, B. F. (2009). The burden of knowledge and the "death of the renaissance man": Is innovation getting harder? *The Review of economic studies*, *76*(1), 283-317. https://doi.org/10.1111/j.1467-937X.2008.00531.x

Leahey, E. (2007). Not by productivity alone: How visibility and specialisation contribute to academic earnings. *American sociological review*, *72*(4), 533-561. https://doi.org/10.1177/000312240707200403

Leahey, E., Keith, B., & Crockett, J. (2010). Specialisation and promotion in an academic discipline. *Research in Social Stratification and Mobility*, *28*(2), 135-155. https://doi.org/10.1016/j.rssm.2009.12.001

Madsen, E. B., & Nielsen, M. W. (2024). Do thematic funding instruments lead researchers in new directions? Strategic funding priorities and topic switching among British grant recipients. *Research Evaluation*, rvae015. https://doi.org/10.1093/reseval/rvae015

Mancuso, R., & Broström, A. (2026). Do mission-oriented grant schemes shape the direction of science? *Research Policy*, *55*(1). https://doi.org/10.1016/j.respol.2025.105360

Myers, K. (2020). The Elasticity of Science. *American Economic Journal-Applied Economics*, *12*(4), 103-134. https://doi.org/10.1257/app.20180518.

Sidorov, G., Gelbukh, A., Gómez-Adorno, H., & Pinto, D. (2014). Soft similarity and soft cosine measure: Similarity of features in vector space model. *Computación y Sistemas*, *18*(3), 491-504. https://doi.org/https://doi.org/10.13053/CyS-18-3-2043.

Stephan, P., Veugelers, R., & Wang, J. (2017). Reviewers are blinkered by bibliometrics. *Nature*, *544*(7651), 411-412. https://doi.org/10.1038/544411a

Tahamtan, I., Safipour Afshar, A., & Ahamdzadeh, K. (2016). Factors affecting number of citations: a comprehensive review of the literature. *Scientometrics*, *107*(3), 1195-1225. https://doi.org/10.1007/s11192-016-1889-2.

Teodoridis, F., Bikard, M., & Vakili, K. (2019). Creativity at the knowledge frontier: The impact of specialisation in fast-and slow-paced domains. *Administrative Science Quarterly*, *64*(4), 894-927. https://doi.org/10.1177/0001839218793384

Torvik, Vetle I., & Smalheiser, Neil R. (2018): Author-ity 2009 - PubMed author name disambiguated dataset. University of Illinois at Urbana-Champaign. https://doi.org/10.13012/B2IDB-4222651_V1.

U.S. National Library of Medicine (2024). *Introduction to MeSH.* https://www.nlm.nih.gov/mesh/introduction.html.

Waltman, L., & Schreiber, M. (2012). On the calculation of percentile-based bibliometric indicators. *Journal of the American Society for Information Science and Technology*, *64*(2), 372-379. https://doi.org/10.1002/asi.22775.

Whitley, R., Glaser, J., & Laudel, G. (2018). The Impact of Changing Funding and Authority Relationships on Scientific Innovations. *Minerva*, *56*(1), 109-134. https://doi.org/10.1007/s11024-018-9343-7

Wuchty, S., Jones, B. F. & Uzzi, B. The increasing dominance of teams in production of knowledge. *Science* 316, 1036–1039 (2007). https://doi.org/10.1126/science.1136099

de Rassenfosse, G., Higham, K., & Penner, O. (2022). Scientific rewards for biomedical specialisation are large and persistent. *Bmc Biology*, 20(1), 211. https://doi.org/10.1186/s12915-022-01400-5

Yu, X., Szymanski, B. K., & Jia, T. (2021). Become a better you: Correlation between the change of research direction and the change of scientific performance. *Journal of Informetrics*, *15*(3). https://doi.org/10.1016/j.joi.2021.101193

Zeng, A., Shen, Z., Zhou, J., Fan, Y., Di, Z., Wang, Y., Stanley, H. E., & Havlin, S. (2019). Increasing trend of scientists to switch between topics. *Nature communications*, 10(1), 3439. https://doi.org/10.1038/s41467-019-11401-8

## 6. APPENDIX

### 6.1. Alternative realisations of the Bernoulli trial

To test the robustness of the results to alternative realisations of the Bernoulli trial used to dichotomize the top 5% indicator, we re-estimate models IV, V, VIII, and IX across 10,000 independent trials. Table A1 shows descriptive statistics for the distribution of coefficient estimates and percent top 5% publications, while figure A1 visualises their distribution. The team and year fixed effects sample only includes publications by teams that have both a top 5% and a non-top 5% publication within the year of the focal publication. This sample consequently has a high proportion of top 5% publications (43%), a higher proportion of publications with a tie citation score, and a larger influence of the Bernoulli trial's randomness. Therefore, using the extremes of this distribution across 10,000 Bernoulli trials represent the hardest robustness test.

**Table A1** Descriptive statistics for the distribution of coefficient estimates and percent top 5% publications. The distribution is created by repeating the Bernoulli trial used to dichotomize the top 5% indicator 10,000 times.

| | | Standard deviation | Min. | Mean | Max |
|---|---|---|---|---|---|
| Full sample | % top 5% | 0.00337 | 6.7906 | 6.80332 | 6.81851 |
| | Specialisation coefficient | 0.00308 | 1.36166 | 1.37462 | 1.38591 |
| | Experience coefficient | 0.00083 | 0.37945 | 0.38256 | 0.38563 |
| Team and year fixed effects | % top 5% | 0.0868 | 43.02793 | 43.40122 | 43.74121 |
| | Specialisation coefficient | 0.02679 | 1.08476 | 1.19263 | 1.29663 |
| | Experience coefficient | 0.00904 | 0.2836 | 0.31543 | 0.35586 |

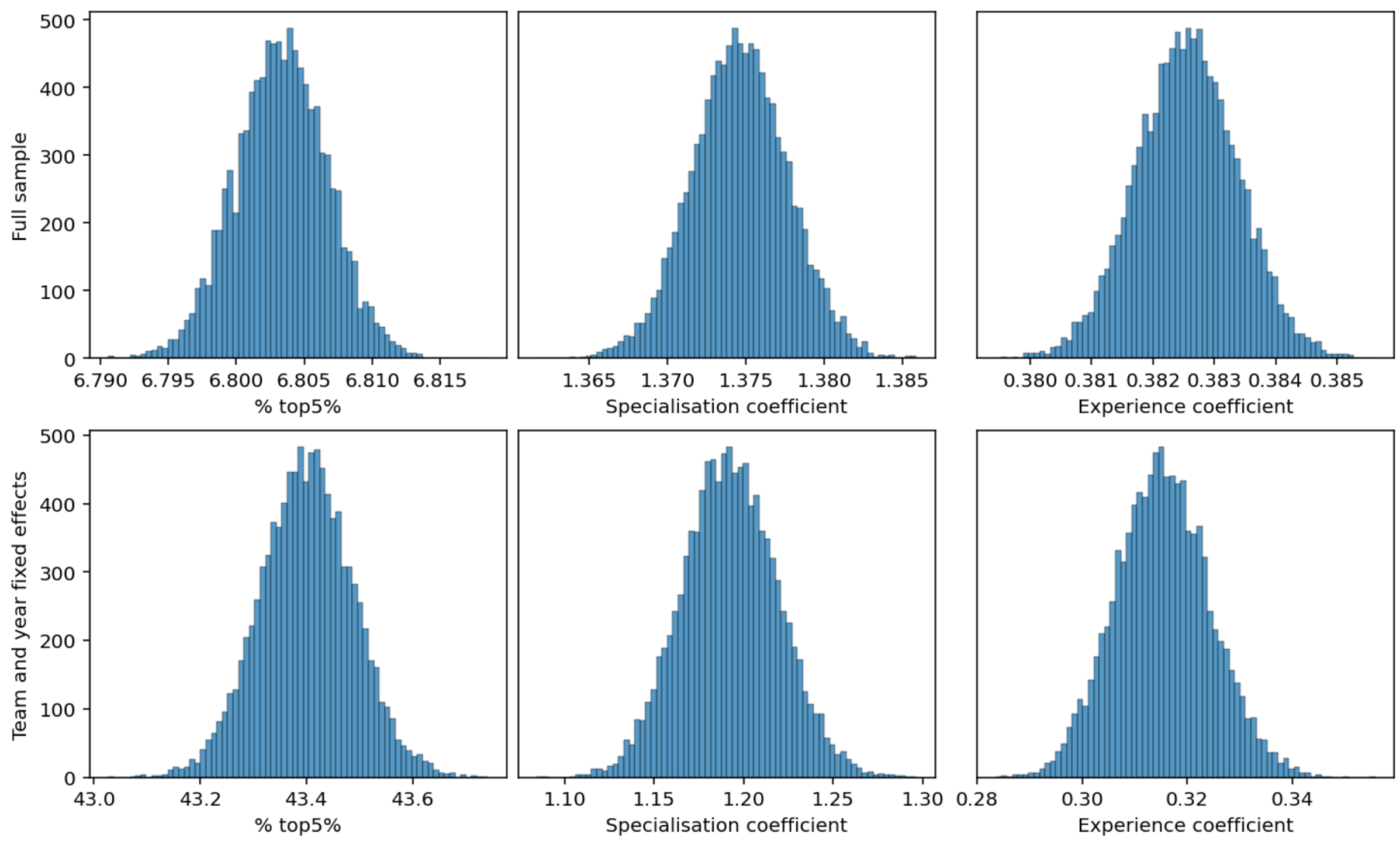

**Fig. A1** Histograms for the distribution of coefficient estimates and percent top 5% publications. The distribution is created by repeating the Bernoulli trial used to dichotomize the top 5% indicator 10,000 times.

We test the robustness of two conclusions from the main text. First, that both specialisation and experience have a substantial correlation with top 5%. This remains true even when using the smallest coefficients obtained out of all 10,000 Bernoulli trials. The difference in odds for a top 5% publication comparing teams one standard deviation above and below the mean of specialisation is 44 % $\left[\exp\left(1.08 \times (0.508 - 0.171)\right) \approx 1.44\right]$. For experience, this difference in odds is 94 % $\left[\exp\left(0.28 \times (5.07 - 2.73)\right) \approx 1.94\right]$. Although these minimum coefficients indicate a weaker association than reported in the main text (55 % and 133 % odds difference, respectively), their magnitudes are still substantial. The second conclusion we test is that the correlation magnitude with top 5% is substantially larger for experience than for specialisation. This also remains true even when comparing the minimum coefficient for experience with the maximum coefficient for specialisation. Using the maximum specialisation coefficient, the difference in odds is 55% $\left[\exp\left(1.30 \times (0.508 - 0.171)\right) \approx 1.55\right]$. This makes the difference in odds 71% larger for experience compared to specialisation $\left[\frac{94}{55} \approx 1.71\right]$.

We therefore conclude that the results are robust to the randomness introduced by the Bernoulli trials. Although the exercise above might seem cumbersome, using a Bernoulli trial to dichotomise the top 5% indicator is arguably more statistically sound than assigning publications to the top 5% when their fraction is 0.5 or greater, and undoubtedly more transparent than simply not communicating how publications with a tie citation score are handled (e.g., Teodoridis et al., 2019).

### 6.2. Alternative samples and definitions of the team publication record vector

Tables A2 and A3 show the results of the logit and conditional logit regressions across alternative samples, specifically when excluding publications with less than three authors and when including single-authored publications, respectively. Tables A4 and A5 show the results when defining the team publication record vector as the maximum and mean of the team members' publication record vectors, respectively.

**Table A2** Logit and conditional logit regressions of top 5%. Excluding publications with less than three authors.

| | Model | | | | | | | | |
|---|---|---|---|---|---|---|---|---|---|
| | Logit | | | | | Conditional logit (Team fixed effects) | | Conditional logit (Team and year fixed effects) | |
| | I | II | III | IV | V | VI | VII | VIII | IX |
| Specialisation | 1.520<br>(0.029) | | | 1.500<br>(0.029) | | 0.957<br>(0.174) | | 1.098<br>(0.284) | |
| ln (Experience) | | 0.322<br>(0.005) | | | 0.414<br>(0.007) | | 0.255<br>(0.052) | | 0.320<br>(0.084) |
| Review | | | 1.471<br>(0.013) | 1.451<br>(0.013) | 1.461<br>(0.013) | 1.190<br>(0.069) | 1.196<br>(0.069) | 0.971<br>(0.111) | 0.978<br>(0.111) |
| ln(#Previous publications) | | | 0.192<br>(0.007) | 0.210<br>(0.007) | -0.167<br>(0.009) | -0.911<br>(0.125) | -1.151<br>(0.140) | | |
| #Authors | | | 0.026<br>(0.003) | 0.016<br>(0.003) | 0.024<br>(0.003) | | | | |
| Constant | -3.252<br>(0.012) | -4.087<br>(0.022) | -4.061<br>(0.032) | -4.645<br>(0.034) | -3.895<br>(0.032) | | | | |
| n | 732,328 | 732,328 | 732,328 | 732,328 | 732,328 | 12,938 | 12,938 | 4,693 | 4,693 |
| n top 5% | 45,799 | 45,799 | 45,799 | 45,799 | 45,799 | 5,320 | 5,320 | 2,142 | 2,142 |
| Log likelihood | -169,909 | -169,009 | -164,882 | -163,577 | -163,270 | -4,338 | -4,341 | -1,592 | -1,592 |

Note: Log-odds coefficients (standard errors). Review is coded ‘review = 1’ and ‘research article = 0’.

**Table A3** Logit and conditional logit regressions of top 5%. Including single-authored publications.

| | Model | | | | | | | | |
|---|---|---|---|---|---|---|---|---|---|
| | Logit | | | | | Conditional logit (Team fixed effects) | | Conditional logit (Team and year fixed effects) | |
| | I | II | III | IV | V | VI | VII | VIII | IX |
| Specialisation | 1.438<br>(0.020) | | | 1.344<br>(0.021) | | 1.002<br>(0.069) | | 1.345<br>(0.132) | |
| ln (Experience) | | 0.207<br>(0.003) | | | 0.365<br>(0.005) | | 0.257<br>(0.019) | | 0.352<br>(0.035) |
| Review | | | 1.483<br>(0.008) | 1.462<br>(0.008) | 1.471<br>(0.008) | 1.246<br>(0.025) | 1.247<br>(0.025) | 1.166<br>(0.047) | 1.169<br>(0.047) |
| ln(#Previous publications) | | | 0.164<br>(0.004) | 0.167<br>(0.004) | -0.157<br>(0.006) | 0.092<br>(0.038) | -0.140<br>(0.043) | | |
| #Authors | | | -0.002<br>(0.002) | -0.007<br>(0.002) | -0.001<br>(0.002) | | | | |
| Constant | -3.115<br>(0.008) | -3.393<br>(0.012) | -3.729<br>(0.017) | -4.197<br>(0.019) | -3.569<br>(0.017) | | | | |
| n | 1,198,575 | 1,198,575 | 1,198,575 | 1,198,575 | 1,198,575 | 62,359 | 62,359 | 16,499 | 16,499 |
| n top 5% | 82,464 | 82,464 | 82,464 | 82,464 | 82,464 | 21,680 | 21,680 | 7,343 | 7,343 |
| Log likelihood | -297,777 | -297,603 | -284,068 | -281,966 | -281,408 | -20,017 | -20,023 | -5,379 | -5,381 |

Note: Log-odds coefficients (standard errors). Review is coded ‘review = 1’ and ‘research article = 0.

**Table A4** Logit and conditional logit regressions of top 5%. Team publication record vector is a function of the max of team members' publication record vectors.

| | Model | | | | | | | | |
|---|---|---|---|---|---|---|---|---|---|
| | Logit | | | | | Conditional logit (Team fixed effects) | | Conditional logit (Team and year fixed effects) | |
| | I | II | III | IV | V | VI | VII | VIII | IX |
| Specialisation | 1.511 (0.024) | | | 1.405 (0.024) | | 1.013 (0.108) | | 1.198 (0.197) | |
| ln (Experience) | | 0.295 (0.004) | | | 0.396 (0.006) | | 0.253 (0.029) | | 0.299 (0.052) |
| Review | | | 1.508 (0.010) | 1.488 (0.010) | 1.500 (0.010) | 1.236 (0.036) | 1.239 (0.036) | 1.091 (0.066) | 1.096 (0.066) |
| ln(#Previous publications) | | | 0.166 (0.005) | 0.171 (0.005) | -0.189 (0.007) | -0.242 (0.065) | -0.433 (0.071) | | |
| #Authors | | | 0.001 (0.003) | 0.006 (0.003) | 0.042 (0.003) | | | | |
| Constant | -3.100 (0.009) | -3.678 (0.015) | -3.760 (0.022) | -4.251 (0.024) | -3.543 (0.023) | | | | |
| n | 992,519 | 992,519 | 992,519 | 992,519 | 992,519 | 31,683 | 31,683 | 9,588 | 9,588 |
| n top 5% | 67,476 | 67,476 | 67,476 | 67,476 | 67,476 | 12,299 | 12,299 | 4,365 | 4,365 |
| Log likelihood | -244,524 | -243,473 | -234,008 | -232,329 | -231,615 | -10,365 | -10,371 | -3,187 | -3,189 |

Note: Log-odds coefficients (standard errors). Review is coded 'review = 1' and 'research article = 0.

**Table A5** Logit and conditional logit regressions of top 5%. Team publication record vector is a function of the mean of team members' publication record vectors.

| | Model | | | | | | | | |
|---|---|---|---|---|---|---|---|---|---|
| | Logit | | | | | Conditional logit (Team fixed effects) | | Conditional logit (Team and year fixed effects) | |
| | I | II | III | IV | V | VI | VII | VIII | IX |
| Specialisation | 1.687<br>(0.025) | | | 1.545<br>(0.026) | | 1.210<br>(0.121) | | 1.432<br>(0.224) | |
| ln (Experience) | | 0.379<br>(0.005) | | | 0.432<br>(0.006) | | 0.288<br>(0.030) | | 0.326<br>(0.055) |
| Review | | | 1.508<br>(0.010) | 1.484<br>(0.010) | 1.490<br>(0.010) | 1.236<br>(0.036) | 1.239<br>(0.036) | 1.092<br>(0.066) | 1.096<br>(0.066) |
| ln(#Previous publications) | | | 0.166<br>(0.005) | 0.169<br>(0.005) | -0.176<br>(0.007) | -0.271<br>(0.066) | -0.450<br>(0.070) | | |
| #Authors | | | 0.001<br>(0.003) | 0.018<br>(0.003) | 0.075<br>(0.003) | | | | |
| Constant | -3.081<br>(0.008) | -3.770<br>(0.015) | -3.760<br>(0.022) | -4.269<br>(0.024) | -3.639<br>(0.022) | | | | |
| n | 992,519 | 992,519 | 992,519 | 992,519 | 992,519 | 31,683 | 31,683 | 9,588 | 9,588 |
| n top 5% | 67,476 | 67,476 | 67,476 | 67,476 | 67,476 | 12,299 | 12,299 | 4,365 | 4,365 |
| Log likelihood | -244,424 | -242,860 | -234,008 | -232,351 | -231,659 | -10,358 | -10,363 | -3,185 | -3,188 |

Note: Log-odds coefficients (standard errors). Review is coded 'review = 1' and 'research article = 0.